# GaMnAs-based magnetic tunnel junctions with an AlMnAs barrier

Shinobu Ohya[a)]

*Department of Electrical Engineering and Information Systems, The University of Tokyo, 7-3-1 Hongo, Bunkyo-ku, Tokyo 113-8656, Japan, and PRESTO Japan Science and Technology Agency, 4-1-8 Honcho, Kawaguchi, Saitama 332-0012, Japan*

Iriya Muneta, Pham Nam Hai, and Masaaki Tanaka[b)]

*Department of Electrical Engineering and Information Systems, The University of Tokyo, 7-3-1 Hongo, Bunkyo-ku, Tokyo 113-8656, Japan*

We investigate the spin-dependent transport of GaMnAs-based magnetic tunnel junctions (MTJs) containing a paramagnetic AlMnAs barrier with various thicknesses. The barrier height of AlMnAs with respect to the Fermi level of GaMnAs is estimated to be 110 meV. We observe tunneling magnetoresistance (TMR) ratios up to 175% (at 2.6 K), which is higher than those of the GaMnAs-based MTJs with other barrier materials in the same temperature region. These high TMR ratios can be mainly attributed to the relatively high crystal quality of AlMnAs and the suppression of the tunneling probability at the in-plane wave-vector $\mathbf{k}_\parallel\sim\mathbf{0}$.

a) Electronic mail: ohya@cryst.t.u-tokyo.ac.jp
b) Electronic mail: masaaki@ee.t.u-tokyo.ac.jp

Coherent transport in III-V-based ferromagnetic-semiconductor GaMnAs heterostructures, which is due to their high-quality epitaxial single crystallinity, is a very attractive feature for future spintronic devices. This will allow effective designing of semiconductor devices using the spin degrees of freedom by utilizing the band-engineering technique. Recently, GaMnAs-based magnetic tunnel junctions (MTJs) with various barrier materials, such as AlAs,[1,2] GaAs,[3] InGaAs,[4] and ZnSe,[5] were studied, and relatively large tunneling magnetoresistance (TMR) values have been obtained. Also, TMR oscillations induced by resonant tunneling have been observed in GaMnAs quantum-well double-barrier heterostructures, proving that holes indeed have high coherency during the tunneling.[6] One of the advantages of the coherent tunneling feature in these structures is that we can find some clues for clarifying the bandstructure of the GaMnAs-based heterostructures from the spin-dependent transport results. Here, we show a systematic study on the GaMnAs-based MTJs with a paramagnetic $Al_{1-y}Mn_yAs$ ($y$=0.05, 0.12) tunnel barrier with various thicknesses from 2 to 5 nm and TMR ratios up to 175% at 2.6 K, which is the highest TMR value ever reported at the same temperature region in the GaMnAs-based MTJs. The mechanism of the TMR enhancement obtained in our study is discussed by analyzing the complex band structure of the tunnel barrier.

Figure 1(a) illustrates the schematic structure of the studied MTJ comprising, from top to bottom, $Ga_{1-x}Mn_xAs$(10 nm)/ GaAs(1 nm)/ $Al_{1-y}Mn_yAs$($d$ nm)/ GaAs($t$ nm)/ $Ga_{1-x}Mn_xAs$(2-10 nm)/ GaAs:Be (Be: $2\times10^{18}$ cm$^{-3}$) grown on a p$^+$GaAs(001) substrate by low-temperature molecular-beam epitaxy (LT-MBE). We grew two series of MTJ samples with the Mn concentration $y$ of $Al_{1-y}Mn_yAs$ set at 5 and 12%. The details of these structures and growth conditions are shown in Table I. After growth, we fabricated round 200-μm-diameter mesa diode structures by standard photolithography and chemical etching. We spin-coated an insulating negative resist on the sample, opened a contact hole with 180 µm diameter on top of the mesa, and fabricated a metal electrode by evaporating Au on this surface. We confirmed that the AlMnAs layer was paramagnetic even at 2.6 K by the magnetic-field dependence of the magnetic circular dichroism (MCD) measurements, which is consistent with the previous studies on AlMnAs films.[7,8] The Curie temperature of the GaMnAs layers of these MTJ samples was around 60 K, which was confirmed by MCD and anomalous Hall effect measurements on the GaMnAs films fabricated with the same growth conditions as those used for the MTJ samples studied here. In the following measurements, the bias polarity is defined by the voltage of the top GaMnAs electrode with respect to the substrate.

Figure 1(b) shows the resistance area ($RA$) product at 3.5 K as a function of the barrier thickness $d$ of $Al_{1-y}Mn_yAs$ obtained in parallel magnetization at zero magnetic field with a bias voltage of 1 mV. As references, experimental $RA$-$d$ data of GaMnAs-based MTJs with AlAs[1] and GaAs tunnel barriers are also shown in Fig. 1(b). Except for the data when the $Al_{1-y}Mn_yAs$ ($y$=5%, 12%) thickness is 2 nm, $RA$ exponentially increases with increasing $d$, which indicates that AlMnAs acts as a tunnel barrier against GaMnAs as well as AlAs and GaAs. We note that the difference of $RA$ between $y$=5% and 12% is due to the difference of the GaAs spacer layer thickness $t$: $t$=1 nm for $y$=5%, and $t$=2 nm for $y$=12%. By fitting to the $RA$-$d$ data with $d$ from 3 to 5 nm using the Wentzel-Kramers-Brillouin (WKB) approximation, the barrier height of $Al_{1-y}Mn_yAs$ with respect to the Fermi level of GaMnAs is estimated to be 110 meV in both cases of $y$=5 and 12%. Here, we assumed that the effective mass of the tunneling hole $m^*$ is the same as that of the heavy hole of AlAs: $0.7m_0$, where $m_0$ is the free electron mass. Similarly, the barrier height of GaAs and AlAs is estimated to be 80 and 450 meV, respectively. Thus, we can find that the barrier height is largely modulated by the Mn doping into AlAs. However, it is still worth paying attention to the deviation from the above fitting curves when the AlMnAs thickness is 2 nm, where the $RA$ values become close to that of the AlAs barrier. This result indicates that the band offset between AlMnAs and GaMnAs is close to that between GaMnAs and AlAs. From these results, we can draw a schematic band diagram of the GaMnAs-based MTJ with the AlMnAs barrier as shown in Fig. 1(c). Here, the GaAs spacer layers are omitted for simplicity. The valence band of AlMnAs is strongly bended near the interfaces, reflecting the large Mn concentration up to ∼$10^{21}$ cm$^{-3}$, where the depletion layer width is estimated to be less than 0.5 nm at each interface. In the middle region of the AlMnAs barrier, the Fermi level is fixed at the energy level 110 meV higher than the valence band. This Fermi level pinning in AlMnAs can be attributed to the various defects in AlMnAs induced by the low-temperature growth.

As an example of the TMR measurements, the inset of Fig. 2(a) shows the magnetic-field $H$ dependence of the $RA(H)$ of the GaMnAs-based MTJ with a 4-nm-thick $Al_{0.95}Mn_{0.05}As$ tunnel barrier at 2.6 K with a magnetic field applied in plane along the [100] axis when the bias voltage is 1 mV. Here, we define the TMR ratio as $[RA_{max}-RA(0)]/RA(0)$, where $RA_{max}$ is the maximum $RA(H)$ in the TMR curve shown in the inset of Fig. 2(a). The red and blue curves were obtained by sweeping the field from -10 to +10 kOe and +10 to -10 kOe, respectively. Typical TMR curves were obtained as shown in the inset of Fig. 2(a),[1,6] and the TMR ratio is estimated to be 175%. The main graph of Fig. 2(a) shows the temperature dependence of the TMR ratio obtained in this device. We can see that the TMR signal persists up to 60 K, which is consistent with our estimation of the Curie temperature of the GaMnAs layers in the MTJs studied here. The TMR values at 2.6 - 5 K shown in Fig. 2(a) are the highest among those ever reported in the GaMnAs-based MTJs in the same temperature region.

Figure 2(b) shows the bias dependence of the TMR of the MTJs with the $Al_{0.88}Mn_{0.12}As$ barrier thickness of 3 nm (red), 4.5 nm (green), and 5 nm (blue). These measurements were carried out at 3.5 K with a magnetic field applied in plane along the [100] axis.  As shown in the inset of the magnified view of Fig. 2(b), the TMR is suppressed in the negative bias region near zero bias in all the data.  Also, in the case of $d$=3 nm, the TMR is strongly suppressed over a wide range of the voltage (~±15 mV) around zero bias.  It is plausible to think that these singular behaviors of the TMR are due to the resonant tunneling through the impurity states[9] of the Mn atoms in the AlMnAs barrier, which are located about 500 meV higher than the valence band top of AlMnAs.[7]  As shown in Fig. 1(c), holes are affected by these impurity levels especially near the interfaces in the AlMnAs barrier because these levels are close to the Fermi level.  When $d$ gets thinner, this effect becomes larger compared to the total tunneling sequence, thus suppressing the TMR near zero bias more strongly.

Figure 3(a) shows the $d$ dependence of the TMR ratio of the MTJs with an AlMnAs barrier.  Because the bias dependence of TMR varies depending on $d$ as shown in Fig. 2(b), the bias voltage used in these measurements was adjusted so that the TMR ratio became maximum in each device.  The TMR ratio was almost saturated when $d$ exceeded 3 nm in both cases of $y$=5% and 12% at 3.5 K.  The TMR suppression in the small $d$ region is presumably due to the magnetic coupling between the GaMnAs layers in the MTJs.

Here, we compare the AlMnAs barrier with the GaAs barrier using their complex bandstructures to investigate the TMR enhancement observed with the AlMnAs barrier. Solid lines in Fig. 3(b) correspond to the simplified valence-band alignment of the GaMnAs-based MTJs used in our calculation.  Assuming that they have parabolic bandstructures and neglecting the thin band-bending regions near the interfaces in the AlMnAs barrier, the tunneling probability at each in-plane wave-vector $\mathbf{k}_\parallel$ is approximately proportional to $\exp\left(-2d\sqrt{2m^*V/\hbar^2 + k_\parallel^2}\right)$, where $V$ is the barrier height, and $k_\parallel$=|$\mathbf{k}_\parallel$|.  Figure 3(c) shows the decay constant $\kappa = \sqrt{2m^*V/\hbar^2 + k_\parallel^2}$ as a function of $k_\parallel$ at 80 meV higher than the valence-band top of GaAs (thin blue curve) and at 110 meV higher than that of AlMnAs (thick red curve), respectively. Here, we neglected the influence of the Mn $d$ states in the AlMnAs barrier for simplicity. The effective mass of tunneling holes in GaAs and AlMnAs was assumed to be $0.45m_0$ and $0.7m_0$, respectively. As shown in the previous study on the GaMnAs-based MTJs with the AlAs barrier, it is important to suppress the tunneling probability in the vicinity of $k_{\parallel}=0$ in order to increase TMR because of the lower spin polarization near $k_{\parallel}=0$ than that in the large $k_{\parallel}$ region.[1,10] When $k_{\parallel}$ is large, $\kappa$ gets closer to the value of $k_{\parallel}$ and becomes less material-dependent with increasing $k_{\parallel}$. On the other hand, near $k_{\parallel}=0$, $\kappa$ is increased by increasing $m^*$ or $V$. Thus, we can see that a tunnel barrier with a large barrier height and a large effective mass is appropriate for obtaining a large TMR. As shown in Fig. 3(c), $\kappa$ at $k_{\parallel}=0$ in AlMnAs is larger than that in GaAs because AlMnAs has a larger barrier height and a larger effective mass than GaAs, which leads to the higher TMR values with the AlMnAs barrier. We also calculated quantitatively their complex band structures assuming more realistic band structures using the nearest-neighbor $sp^3s^*$ tight-binding model[11,12] and obtained the same tendency described above. The details will be published elsewhere.

Finally, we compare the AlAs barrier with the AlMnAs barrier. Although AlAs has a higher barrier height than AlMnAs, the TMR of MTJs with an AlAs barrier is not so large as that with an AlMnAs barrier. This can be attributed to the worse crystal quality of AlAs. During the MBE growth, the reflection high energy electron diffraction (RHEED) of AlMnAs was a more streaky 1×1 pattern than that of AlAs when the substrate temperature was below 200ºC, which led to the higher TMR ratios when the AlMnAs barrier was used. If we can improve the crystal quality of the low-temperature grown AlAs tunnel barrier, a very large TMR will be obtained. Another possibility is that the TMR is enhanced by magnetic-impurity assisted tunneling.[13] In this case, the interstitial and substitutional Mn defects in the AlMnAs barrier can enhance the TMR, though this effect may be small because these Mn impurity states are far from $E_F$ in almost all the AlMnAs barrier region (See Fig. 1(c)).

In summary, we investigated the spin-dependent tunneling in the GaMnAs-based MTJs containing a paramagnetic $Al_{1-y}Mn_yAs$ ($y$=0.05, 0.12) tunnel barrier with various thicknesses in the range $d$ = 2-5 nm. The barrier height of AlMnAs with respect to the Fermi level of GaMnAs was estimated to be 110 meV if the effective mass of the tunneling carrier was assumed to be the same as that of the heavy hole in AlAs. The TMR ratio increased with increasing $d$, but saturated at $d > 3$ nm at 3.5 K. The TMR ratio reached up to 175% for $d$=4 nm at 2.6 K. The TMR enhancement observed in our MTJ devices can be mainly attributed to the relatively high crystal quality of AlMnAs and the suppression of the tunneling probability at $k_{\parallel} \sim 0$.

This work was partly supported by Grant-in-Aids for Scientific Research, the Special Coordination Programs for Promoting Science and Technology, R&D for Next-generation Information Technology by MEXT, and PRESTO of JST.

TABLE I. Details of the structures and growth conditions of the studied MTJs comprising $Ga_{1-x}Mn_xAs$(10 nm)/ GaAs(1 nm)/ $Al_{1-y}Mn_yAs$($d$ nm)/ GaAs($t$ nm)/ $Ga_{1-x}Mn_xAs$(2-10 nm)/ GaAs:Be on $p^+$GaAs(001) substrates. "Bottom" and "Top" mean the bottom and top GaMnAs layers, respectively.

| $y$ (%) | $x$ (%) | $d$ (nm) | $t$ (nm) | Growth temperature (ºC) | | |
| --- | --- | --- | --- | --- | --- | --- |
| | | | | Bottom | GaAs/AlMnAs/GaAs | Top |
| 5 | 5 | 2-5 | 1 | 225 | 200 | 225 |
| 12 | 6 | 2-5 | 2 | 215 | 170 | 190 |

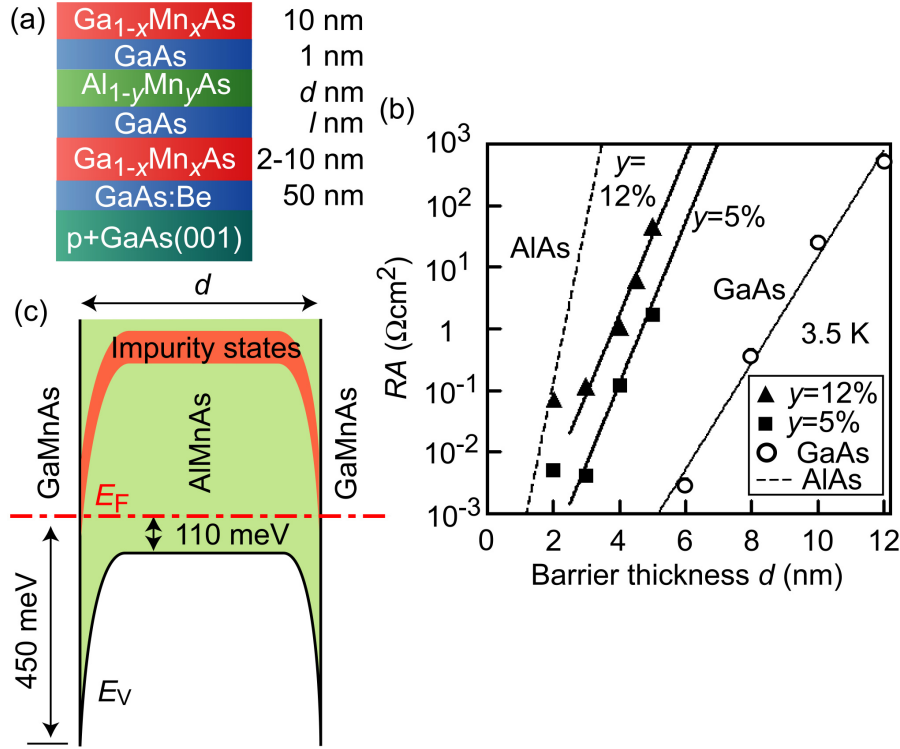

FIG. 1. (a) Schematic GaMnAs-based MTJ structure with an AlMnAs barrier studied here. (b) Resistance area ($RA$) product as a function of the barrier thickness in the GaMnAs-based MTJs when the tunnel barrier is AlAs, $Al_{1-y}Mn_yAs$ ($y$=5%, 12%), and GaAs. These data were obtained at zero magnetic field in parallel magnetization with a bias voltage of 1 mV at 3.5 K. The solid rectangles and solid triangles are the data of MTJs with an $Al_{1-y}Mn_yAs$ barrier with $y$=5% and 12%, respectively. The broken line and the open circles correspond to the data of the GaMnAs-based MTJs with AlAs[1] and GaAs tunnel barriers, respectively. (c) Schematic valence-band diagram of the GaMnAs-based MTJ with an AlMnAs barrier. Here, the GaAs spacer layers are omitted for simplicity.

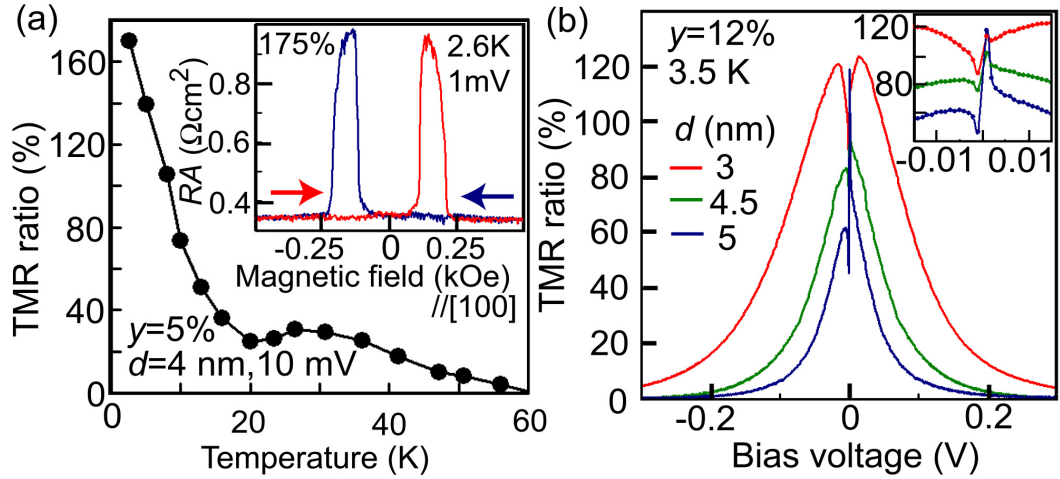

FIG. 2. (a) Inset shows the magnetic-field dependence of $RA$ of the MTJ with a 4-nm-thick $Al_{0.95}Mn_{0.05}As$ barrier at 2.6 K with a magnetic field applied in plane along the [100] axis when the bias voltage is 1 mV. The main graph shows the temperature dependence of TMR of this device with a magnetic field applied in plane along the [100] axis when the bias voltage is 10 mV. (b) Bias dependence of TMR of the GaMnAs-based MTJs with the $Al_{0.88}Mn_{0.12}As$ thickness of 3 nm (red), 4.5 nm (green), and 5 nm (blue) when the magnetic field is applied in plane along the [100] axis. The inset shows the magnified view near zero bias.

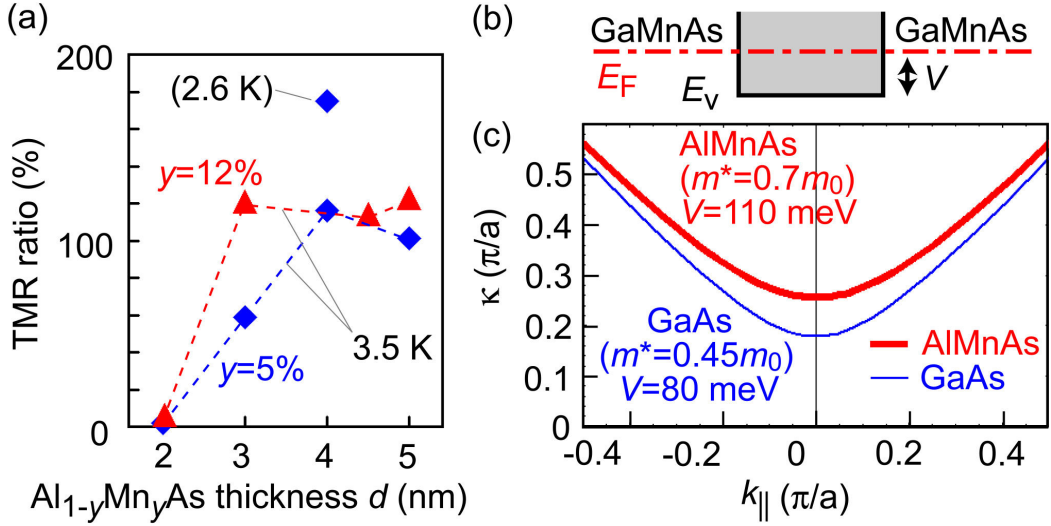

FIG. 3. (a) $Al_{1-y}Mn_yAs$ barrier thickness $d$ dependence of the TMR, where the bias voltage was set so that the TMR became maximum in each device due to the peculiar bias dependence as shown in Fig. 2(b). The rhombic and triangle points correspond to $y$=5% and 12%, respectively. The measurement temperature was 3.5 K, except for the data of 175% at $y$=5% and $d$=4 nm measured at 2.6 K. (b) Simplified band diagram of the MTJ structure used for our calculation of the decay constant $\kappa$. (c) Calculated $\kappa$ as a function of $k_\parallel$ at 80 meV higher than the valence-band top of GaAs (thin curve) and at 110 meV higher than that of AlMnAs (thick curve), respectively. Here, the effective mass $m^*$ in GaAs and AlMnAs was assumed to be $0.45m_0$ and $0.7m_0$, respectively.